# Critical change in the Fermi surface of iron arsenic superconductors at the onset of superconductivity

Chang Liu[1*], Takeshi Kondo[1*], Rafael M. Fernandes[1], Ari D. Palczewski[1], Eun Deok Mun[1], Ni Ni[1], Alexander N. Thaler[1], Aaron Bostwick[2], Eli Rotenberg[2], Jörg Schmalian[1], Sergey L. Bud'ko[1], Paul C. Canfield[1], and Adam Kaminski[1]

(Dated: September 18, 2009)

PACS: 79.60.-i, 74.25.Jb, 74.70.Dd

[1] Ames Laboratory and Department of Physics and Astronomy, Iowa State University, Ames, Iowa 50011, USA
[2] Advanced Light Source, Lawrence Berkeley National Laboratory, Berkeley, California 94720, USA
* These authors contribute equally to this work.

The phase diagram of a correlated material is the result of a complex interplay between several degrees of freedom, providing a map of the material's behavior. One can understand (and ultimately control) the material's ground state by associating features and regions of the phase diagram, with specific physical events or underlying quantum mechanical properties. The phase diagram of the newly discovered iron arsenic high temperature superconductors [1,2] is particularly rich and interesting. In the $AE(Fe_{1-x}T_x)_2As_2$ class ($AE$ being Ca, Sr, Ba, $T$ being transition metals), the simultaneous structural/magnetic phase transition that occurs at elevated temperature in the undoped material, splits and is suppressed by carrier doping, the suppression being complete around optimal doping[3,4,5,6]. A dome of superconductivity exists with apparent equal ease in the orthorhombic / antiferromagnetic (AFM) state as well as in the tetragonal state with no long range magnetic order[3,7,8,9,10]. The question then is what determines the critical doping at which superconductivity emerges, if the AFM order is fully suppressed only at higher doping values. Here we report evidence from angle resolved photoemission spectroscopy (ARPES) that critical changes in the Fermi surface (FS) occur at the doping level that marks the onset of superconductivity. The presence of the AFM order leads to a reconstruction of the electronic structure, most significantly the appearance of the small hole pockets at the Fermi level. These hole pockets vanish, i. e. undergo a Lifshitz transition, at the onset of superconductivity. Superconductivity and magnetism are competing states in the iron arsenic superconductors[7]. In the presence of the hole pockets superconductivity is fully suppressed, while in their absence the two

**states can coexist.**

Previous studies demonstrated that the presence of the long range AFM order in the undoped pnictides leads to a substantial reconstruction of the Fermi surface[11,12,13,14,15,16]. In Figs. 1a, 1b we present the FS maps of $Ba(Fe_{1-x}Co_x)_2As_2$ at $T =$ 20K for two extreme doping levels: $x = 0$ where AFM order is present, and $x = 0.114$ where it is fully suppressed by cobalt doping. In the undoped samples the $X$-pocket FS looks like four flower petals[15] – high intensity peaks are visible along the diagonal $k_{(110)}$ and $k_{(1,-1,0)}$ directions. For heavily overdoped, paramagnetic samples ($x = 0.114$), the $X$-pocket FS changes to an oval shape and the peaks along the $k_{(110)}$ direction are absent. In Figs. 1c - 1j we point out the magnetic origin of this FS reconstruction by comparing the details of the $X$-pocket for the two doping levels and a prediction of a five band tight binding model calculation (similar to Ref. 15) with and without an AFM order. Figs. 1c - 1f show the ARPES intensity plots at the chemical potential ($\mu$) and 50meV below $\mu$ in the vicinity of the $X$-point. Figs. 1g - 1j show the corresponding theoretical prediction. The presence of the AFM order is measured by the mean field order parameter $\Delta_{AF}$, or in other words, the gap opened in the AFM state. The comparison shown in Fig. 1 clearly demonstrates that the observed FS reconstruction is consistent with the effects of a long range AFM order on the electronic structure. In the AFM state, the four Fermi peaks that make up the "flower petals" appear in both the experimental data and the theoretical calculation, and they are *hole-like* – increasing the binding energy results in a larger size of the "petals" in the constant energy cuts. In the PM state, however, these "petals" are absent in both the experiment and theory. The main *X*-pocket is *electron-like* – increasing the binding energy results in smaller pockets. We will utilize these features to quantify the effects of the AFM order on the electronic structure of the pnictides.

In Fig. 2 we plot the ARPES FS maps at $T$ = 13 and 150K for several different cobalt doping levels obtained with a photon energy of 21.2eV (He-I line). It is clear that the reconstructed hole pockets are present for $x < 0.034$ at low temperatures and vanish above the magnetic transition temperature $T^*$. It is worth noting that the temperature dependence of the FS is not due to thermal broadening, since the low and high temperature data for higher doping levels are very similar. The first and most important observation in Fig. 2 is that the FS reconstruction vanishes rather rapidly at intermediate doping levels. The intensity of those reconstructed pieces of the FS at low temperature starts to decrease at a doping level of $x = 0.024$, and effectively the intensity has vanished by $x = 0.038$. This is due to the fact that the top of the hole bands move below the Fermi energy, which is a classic signature of a Lifshitz transition. This occurs at the doping where superconductivity emerges in the phase diagram. In Fig. 2b we quantify this transition by plotting the maximum intensity around the *X*-pocket FS (from panels in Fig. 2a) as a function of the angle ($\alpha$) with respect to the $k_{(110)}$ direction. With hole pockets present ($0 < x < 0.028$) there is a strong peak at $\alpha = 0°$. This peak decreases as *x* changes from 0.024 to 0.034 and vanishes by $x = 0.038$. Beyond this doping these small reconstructed pieces of FS are absent. Neutron scattering experiments[7] clearly demonstrate that long range AFM order coexists and competes with superconductivity for $x = 0.047$, and likely extends up to $x \sim 0.06$. Indeed in ARPES data such reconstruction is observed up to $x = 0.058$, but only at higher binding energies (see supplementary information). This suggests that the response of the FS to the AFM order is essential for the existence of superconductivity. As the doping increases beyond $x \sim 0.03$, the small hole pockets disappear and superconductivity can take place even in the AFM state.

In Fig. 2c we focus on the doping evolution of the nesting condition by examining the contours of the $X$- and $\Gamma$- pocket FS sheets in the doping region where the FS reconstruction is absent. For doping levels in the range $0.038 < x < 0.073$, both pockets are roughly similar in shape and size indicating reasonably good nesting conditions. For heavily overdoped samples, however, the $X$-pocket is significantly larger than the $\Gamma$-pocket. This is noteworthy since this sample is still superconducting ($T_c = 12.8K$) and nesting is considered very important for the pairing mechanism in these materials[17,18,19].

We now compare our ARPES data to recent results of the Hall Effect and thermoelectric power (TEP) measurements[20]. In Fig. 3a we plot the Hall coefficient $R_H$ as a function of doping at $T = 25$ and $150K$[20] measured on samples from the same batch as those used for the ARPES measurements, compared with the maximum ARPES intensity of the reconstructed FS from Fig. 2b (defined in the caption). One can see that the onset of superconductivity $0.024 < x_{CRIT} < 0.034$ correlates with a dramatic change of the Hall coefficient that ends rapidly inside the superconducting dome. In a similar manner the TEP data[20] (not shown) changes abruptly with doping between $x = 0.02$ and $x = 0.024$ for a surprisingly wide range of temperatures (25K < $T$ < 300K). This implies that the TEP data is much more sensitive to the onset of this change – consistent with TEP being more closely related to the derivative of the density of state at the chemical potential with respect to energy (d$N(E_F)$/d$E$) rather than $N(E_F)$ itself. It is possible that finer ARPES measurements can identify the changes in the electronic properties at high temperature observed by TEP.

We summarize our main findings in Fig. 3b, where we plot the locations of each ARPES data on the $x$-$T$ phase diagram. Fig. 3b shows that the emergence of superconductivity in Ba(Fe$_{1-x}$Co$_x$)$_2$As$_2$ coincides with the disappearance of the reconstructed pieces of the Fermi surface. Our observation explains that the changes in Hall coefficient, namely the rapid increase of $R_H$, are caused by critical changes of the Fermi surface topology. The pairing interaction on the reconstructed Fermi surface is reduced due to the requirement that the quasi-particle induced spin wave damping must vanishes at the ordering vector[21]. In the context of the pnictides, this effect leads to the reduced pairing interaction of the magnetically ordered state found by Parker et al.[22]. The modified wave functions in the magnetically ordered state couple less efficiently to magnetic fluctuations. The effect is strongest for large magnetization with pronounced down folding and demonstrates the sensitivity of an electronic pairing mechanism with respect to an antiferromagnetic FS reconstruction. The finding of the present paper clearly demonstrates that avoiding the Fermi surface reconstruction is key to establishing the superconductivity in iron arsenic high temperature superconductors.

**Methods**

Single crystals of Ba(Fe$_{1-x}$Co$_x$)$_2$As$_2$ were grown via self-flux using conventional high-temperature solution growth techniques. The doping level $x$ was determined using wavelength dispersive X-ray spectroscopy in a JEOL JXA-8200 electron microprobe[3]. Transport and magnetization measurements report a tetragonal to orthorhombic structural transition with a paramagnetic to AFM transition at $T^* \approx T_N \approx$ 135K for the undoped ($x = 0$) samples. Increasing the doping significantly suppresses and splits the two transition temperatures[3,7]; superconductivity appears around $x_{CRIT} \sim$ 0.03 and a maximum onset $T_c$ of ~24K was observed for the $x = 0.058$ samples[3]. The ARPES measurements were performed at a laboratory-based ARPES system consisting of a Scienta SES2002 electron analyzer, GammaData UV lamp and custom designed refocusing optics at Ames Laboratory, as well as beamline 7.0.1 of the Advanced Light Source (ALS), Berkeley, California with a Scienta R4000 electron analyzer. Vacuum conditions were better than $3 \times 10^{-11}$ torr. Energy resolution was set at ~25meV for Figs. 1 and ~15meV for Fig. 2. All samples were cleaved *in situ* yielding mirror-like, clean *a-b* surfaces. Cleaved surfaces of all samples were stable for at least 24 hours. Results were reproduced at the Advanced Light Source beamline 7.0.1 as well as Ames Laboratory on several samples. The high symmetry point $X$ is defined to be ($\pm\pi/a$, $\pm\pi/a(b)$, 0) with the $k_x$ and $k_y$ axes along the Fe-As bonds.

# Acknowledgements

We acknowledge useful discussions with J. Schmalian and Y. S. Kim for his excellent instrumental support at ALS. Ames Laboratory is supported by the Department of Energy - Basic Energy Sciences under Contract No. DE-AC02-07CH11358. ALS is operated by the US DOE under Contract No. DE-AC03-76SF00098.

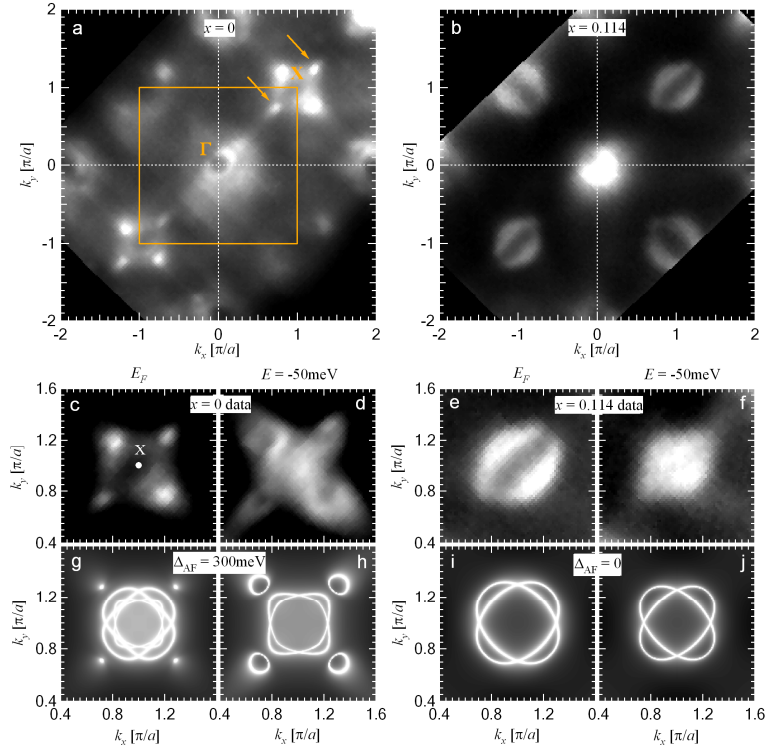

**Figure 1. The Fermi surface reconstruction and its magnetic origin. a-b**, Fermi surface mappings of Ba(Fe$_{1-x}$Co$_x$)$_2$As$_2$ - intensity of the photoelectrons integrated over 10 meV about the chemical potential. Data is taken with 105eV photons in the antiferromagnetic ($x = 0$) and paramagnetic ($x = 0.114$) phases at $T = 20$K. Bright areas indicate higher intensity. Orange arrows emphasize the Fermi peaks along the $k_{(110)}$ direction which are absent in the paramagnetic phase. **c-f**, Expanded ARPES intensity plots of panels **a** and **b** in the vicinity of the *X*-points for two different binding energies indicated at the top of each graph. **g-j**, Results of a 5 band tight binding model calculation for the same binding energies. $\Delta_{AF}$ is a measure of the antiferromagnetic order parameter with unit eV; $\Delta_{AF} = 0$ describes the paramagnetic phase.

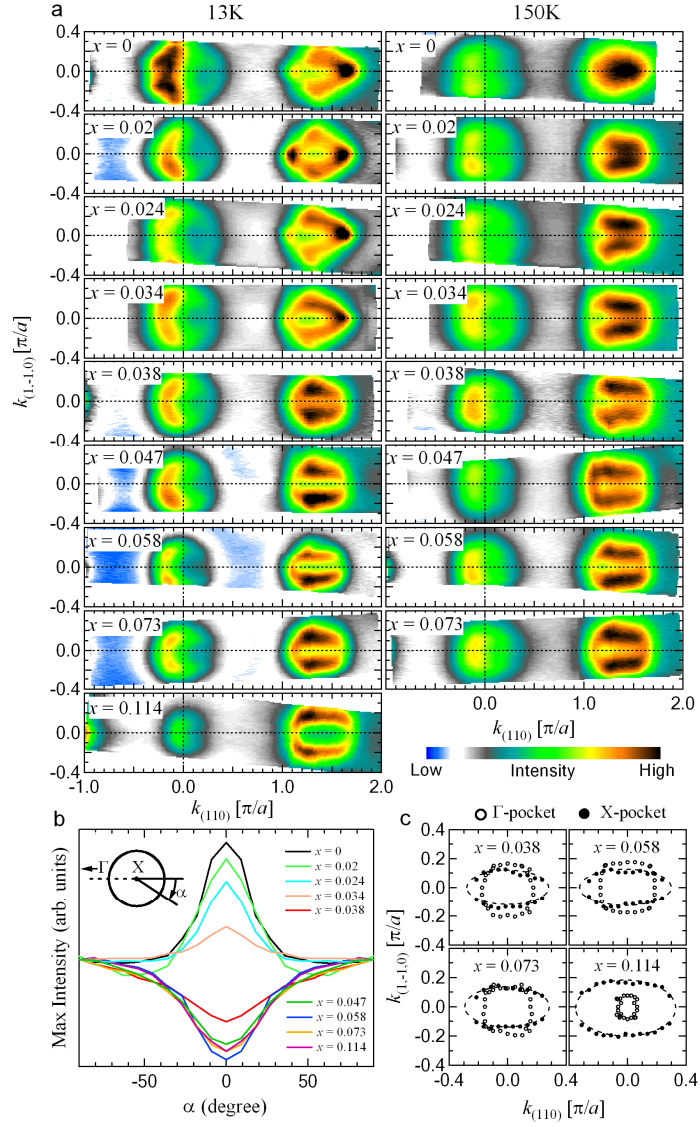

**Figure 2. The vanishing of the dramatic Fermi surface reconstruction coincides with the onset of superconductivity. a**, Fermi surface mappings of Ba(Fe$_{1-x}$Co$_x$)$_2$As$_2$ for temperatures $T = 13$ and 150K measured at various cobalt doping levels. The incident photon energy is 21.2eV. (Note: image slice rotated by 45° with respect to Fig. 1) **b**, Maximum intensity of electrons around the $X$-pocket for the low temperature data, shown as a function of an $\alpha$ angle defined as the angular deviation with respect to the $k_{(110)}$ direction (inset). The intensities are normalized at $\alpha = 90°$ and symmetrized with respect to $\alpha = 0°$. **c**, $\Gamma$- and $X$-pocket location of the low temperature data extracted via the peak position of the momentum distribution curves for $0.038 \leq x \leq 0.114$. The $X$-pocket is shifted to the $\Gamma$-pocket by rescaling the $k_{(110)}$-axis via $k' = k - \sqrt{2}$ for easier comparison of their areas.

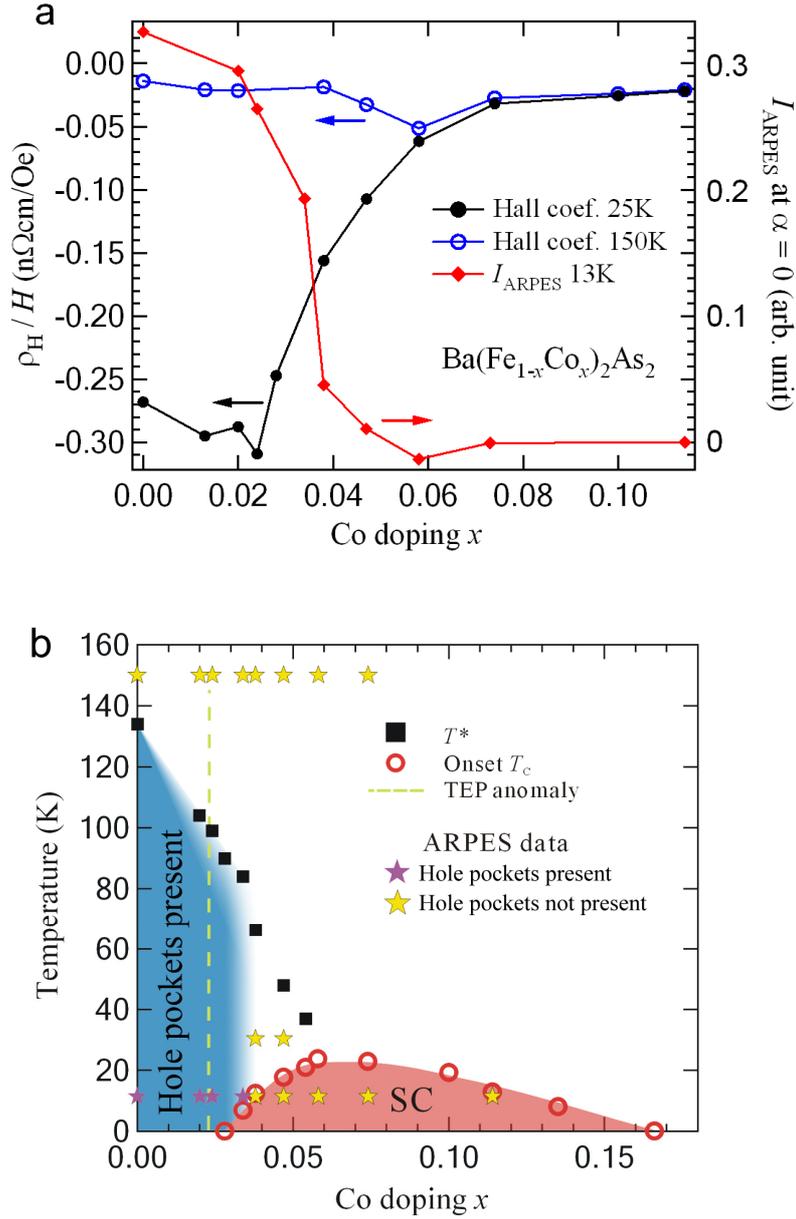

**Figure 3. a**, Left axis: The *a-b* plane Hall coefficient $R_H \equiv \rho_H / H$ vs. $x$ data of Ba(Fe$_{1-x}$Co$_x$)$_2$As$_2$ with magnetic field $H = 90$ kOe$\|c$, measured at two temperatures $T = 25$ and 150K; Data from Ref. 20 is used. Right axis: ARPES intensity $I_{ARPES}$ at $\alpha = 0°$ extracted from Fig. 2b, subtracted by the intensity at the highest doping $x = 0.114$. **b**, Schematic phase diagram of Ba(Fe$_{1-x}$Co$_x$)$_2$As$_2$ based on ARPES and transport measurements. Black squares show the magnetic transition temperature $T^*$ determined by resistivity measurements, red hollow circles show the onset temperature of superconductivity determined by resistivity measurements; Data from Ref. 3 is used. Green dashed line shows the doping location of the sudden change in thermoelectric power; Data from Ref. 20 is used. Blue area demonstrates the doping-temperature region where the reconstructed Fermi surface is present. Red area indicates the superconducting dome. Purple (yellow) stars mark the phase diagram locations of each ARPES intensity map showing a reconstructed (non-reconstructed) Fermi surface. Some data is shown in the online supplementary information.

# Online supplementary information for "Critical change in the Fermi surface of iron arsenic superconductors at the onset of superconductivity"

Chang Liu[1*], Takeshi Kondo[1*], Rafael M. Fernandes[1], Ari D. Palczewski[1],

Eun Deok Mun[1], Ni Ni[1], Alexander N. Thaler[1], Aaron Bostwick[2], Eli Rotenberg[2],

Jörg Schmalian[1], Sergey L. Bud'ko[1], Paul C. Canfield[1] and Adam Kaminski[1]

(Dated: September 18, 2009)

PACS: 79.60.-i, 74.25.Jb, 74.70.Dd

[1] Ames Laboratory and Department of Physics and Astronomy, Iowa State University, Ames, Iowa 50011, USA
[2] Advanced Light Source, Lawrence Berkeley National Laboratory, Berkeley, California 94720, USA
* These authors contribute equally to this work.

1. **Temperature and doping dependence of the Fermi surface in $Ba(Fe_{1-x}Co_x)_2As_2$ measured with 105eV photons.**

ARPES data on the Fermi surface (FS) evolution similar to Fig. 2 of the paper "Critical change in the Fermi surface of iron arsenic superconductors at the onset of superconductivity" is also collected at a higher photon energy which corresponds to a different value of $k_z$. Fig. S1 shows the FS mappings of $Ba(Fe_{1-x}Co_x)_2As_2$ measured via a monochromatic synchrotron photon beam with photon energy $hv$ = 105eV. With a choice of the inner potential $V_0$ = 15eV[1], the $k_z$ value for 105eV and 21.2eV photons differs by $10.9\pi$ (or $1.1\pi$ in the reduced zone scheme) at Γ. It is a notable fact that no apparent difference of the FS reconstruction is observed for these two different $k_z$s. "Petals" and ovals are seen for antiferromagnetic (AFM) and paramagnetic states, respectively, and the FS reconstruction occurs below the doping level $x$ = 0.038. This fact is better seen in Fig. S1(j) where the momentum distribution curves (MDCs) along the $k_{(110)}$ direction are extracted for the low temperature data for the five doping levels. The sharp peaks located at $k_{(110)}$ ~ 1.05 and 1.75 $\pi/a$ disappear when the doping is changed from $x$ = 0.02 to $x$ = 0.038. Both the low and high temperature data is fully consistent with Fig. 2 in the main text, proving that the main conclusion of the paper is reproducible.

2. **Cobalt doping and binding energy dependence of the electronic structure in $Ba(Fe_{1-x}Co_x)_2As_2$.**

Fig. S2 shows the low temperature ($T = 20$K) electronic structure of Ba(Fe$_{1-x}$Co$_x$)$_2$As$_2$ for different Co doping levels and binding energies. From Fig. S2 it is clear that the electronic structure reconstruction due to the presence of the magnetic order is seen up to the doping level of $x = 0.058$, where antiferromagnetism is coexisting with superconductivity. Such a reconstruction presents itself as a typical band back-folding effect, forming hole pockets around the *X*-point at higher binding energies. Therefore the ARPES intensity maps for the $x = 0$ and $x = 0.058$ samples are very similar at binding energies $E_b > 60$meV. What is important is the fact that this effect is absent *at the Fermi level* for the superconducting samples. Typically for $x = 0.058$ ($T_c \sim 24$K), the bright peaks at $E_F$ along the $k_{(110)}$ and $k_{(1,-1,0)}$ directions are absent, in sharp contrast with The FS topology for the undoped sample. This observation fully verifies the main argument of the paper that maintaining the paramagnetic-like electronic structure *at the Fermi level* at low temperatures is a critical ingredient for superconductivity in the iron pnictides.

3. **Temperature dependence of the electronic structure of Ba(Fe$_{0.953}$Co$_{0.047}$)$_2$As$_2$.**

In order to verify the fact that the doping-temperature region for the reconstructed FS follows a straight line up to *T\** around the onset of superconductivity, ARPES intensity map is taken at several temperatures for Ba(Fe$_{0.953}$Co$_{0.047}$)$_2$As$_2$. At this doping level, neutron diffraction data[2] from the *same* growth batch clearly shows three main transitions: a tetragonal to orthorhombic transition at $T_S = 60$K, a second-order magnetic transition from paramagnetic to AFM order at $T^* = 47$K, and a superconducting transition at $T_c = 17$K below which the AFM order coexists with superconductivity. If the vanishing of the electronic structure reconstruction at $E_F$ were solely resulting from the appearance of superconductivity, the AFM reconstructed FS would be expected for $T_c < T < T^*$. Fig. S3 shows the ARPES intensity plot for Ba(Fe$_{0.953}$Co$_{0.047}$)$_2$As$_2$ at $T = 13$, 30 and 150K, among which $T = 30$K lies between $T_c$ and $T^*$. It is very clear from the figure that no observable reconstruction is present close to $E_F$ even at $T = 30$K. This result is not due to aging effect, since the sample was cleaved in the vacuum chamber at 30K and the same ARPES map is measured for several times in a time span of about 48 hours (not shown), no major difference with time is seen. This data proves that there is no FS reconstruction at $x = 0.047$, even though AFM order is confirmed below $T^* = 47$K. Similar temperature dependence data is also taken at $x = 0.038$ (not shown), where again no FS reconstruction can be seen. Therefore, the electronic response to the AFM order at the Fermi level does not coincide with the AFM transition. The pronounced FS reconstruction is absent for all samples beyond the doping level of $x = 0.038$.

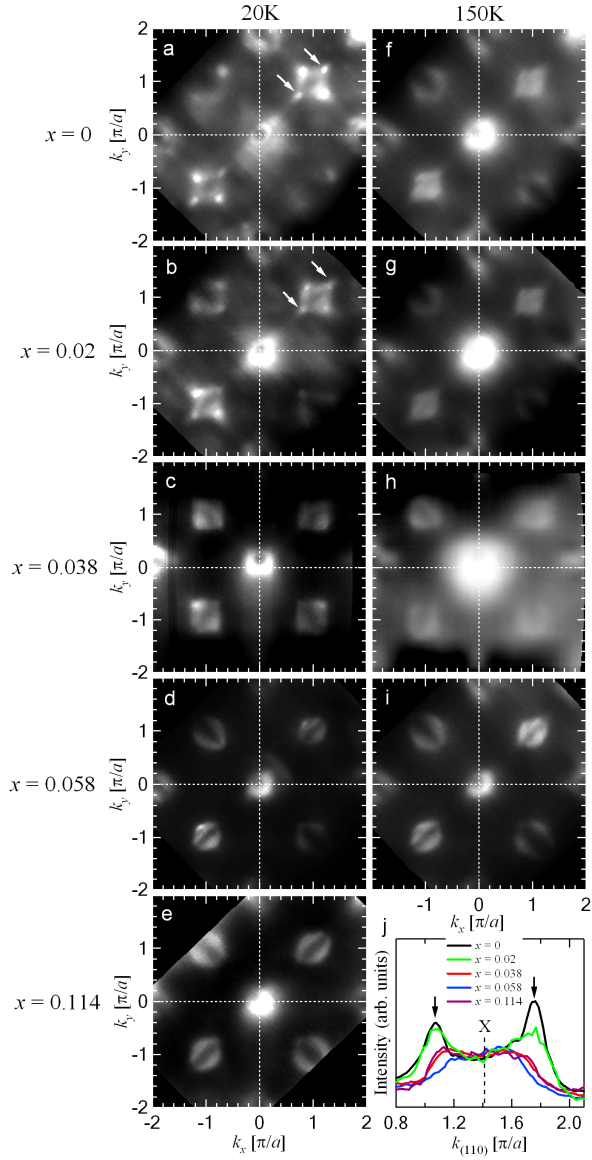

**Figure S1**. **Temperature and doping dependence of the Fermi surface in Ba(Fe$_{1-x}$Co$_x$)$_2$As$_2$ measured with 105eV photons. a-i**, Fermi surface mappings of Ba(Fe$_{1-x}$Co$_x$)$_2$As$_2$ for temperatures $T$ = 20 and 150K measured at various cobalt doping levels. Bright areas indicate higher intensity. White arrows in **a** and **b** emphasize the Fermi peaks along the $k_{(110)}$ direction. Panels **a** and **e** are the same as Fig. 1a and 1b in the main text. **j**, Doping evolution of the momentum distribution curves (MDCs) at the chemical potential along $k_{(110)}$ direction for $T$ = 20K data.

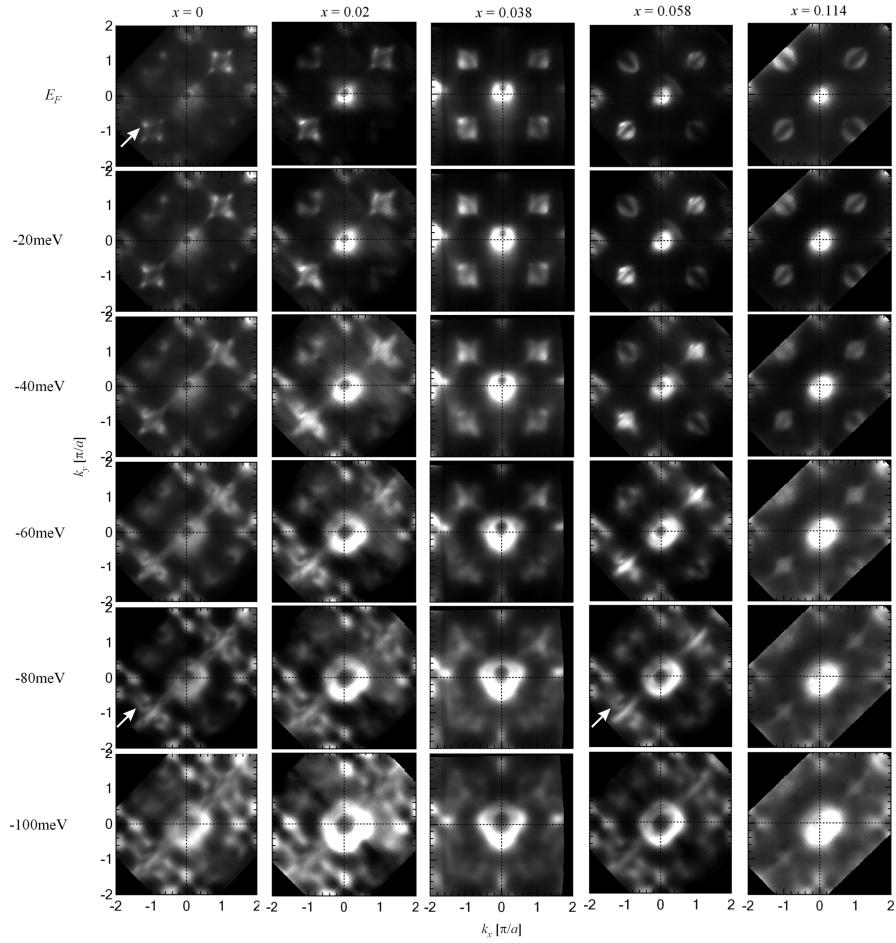

**Figure S2. Doping and binding energy dependence of the electronic structure in Ba(Fe$_{1-x}$Co$_x$)$_2$As$_2$.** Data is taken with 105eV photons at $T$ = 20K. Doping values are indicated at the top of each column and energy values with respect to $E_F$ are shown to the left of each row. Note that at the doping level of $x$ = 0.058 where superconductivity is present with the antiferromagnetic order, the electronic structure reconstruction due to the AFM order is present at higher binding energies ($E_b$ > 60meV) but not at the Fermi level. This fact is indicated by white arrows in the figure.

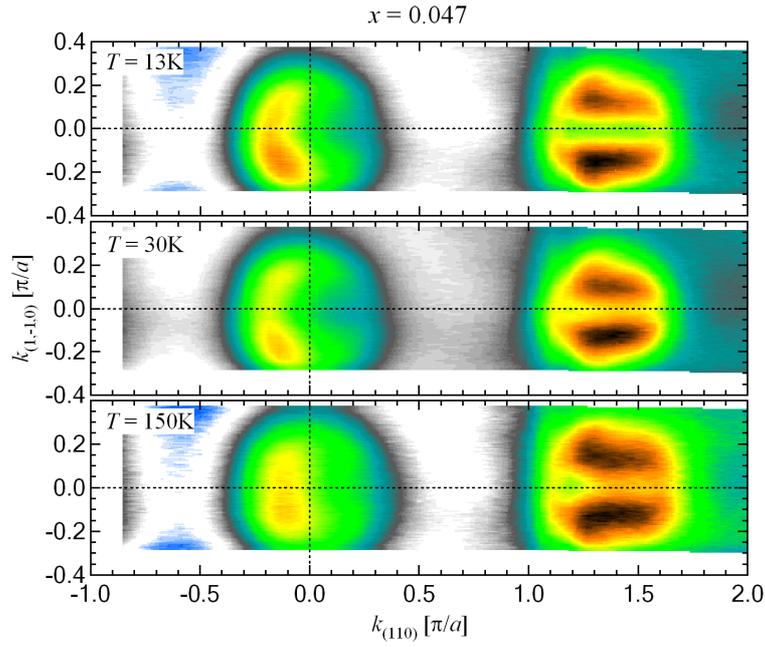

**Figure S3. The electronic structure of Ba(Fe$_{0.953}$Co$_{0.047}$)$_2$As$_2$ at $T$ = 13, 30 and 150K -** Intensity of the photoelectrons integrated over 10 meV about the chemical potential. The incident photon energy is 21.2eV, color scale is the same as that used in Fig. 2 of the main text. The upper panel is the same as the corresponding panel in Fig. 2, the lower panel is taken from a sample different than that used in Fig. 2. Clearly no Fermi surface reconstruction is visible at this doping level.